\documentclass[prb,showpacs,twocolumn,nofootinbib]{revtex4}

\usepackage{amsmath,amssymb,epsfig,epsf}

\begin{document}

\title
{\bf Two--dimensional Anderson--Hubbard model in DMFT+$\Sigma$ approximation. }
\author{E.Z. Kuchinskii, N.A. Kuleeva, I.A. Nekrasov, M.V. Sadovskii}

\affiliation
{Institute for Electrophysics, Russian Academy of Sciences,
Ekaterinburg, 620016, Russia}

\begin{abstract}

Density of states, dynamic (optical) conductivity and phase diagram of
paramagnetic two -- dimensional Anderson -- Hubbard model with strong 
correlations and disorder are analyzed within the generalized dynamical mean
field theory (DMFT+$\Sigma$ approximation). Strong correlations are accounted
by DMFT, while disorder is taken into account via the appropriate generalization
of the self -- consistent theory of localization. We consider the 
two -- dimensional system with the rectangular ``bare'' density of states (DOS).
The DMFT effective single impurity problem is solved by numerical renormalization
group (NRG). Phases of ``correlated metal'', Mott insulator and correlated
Anderson insulator are identified from the evolution of density of states,
optical conductivity and localization length, demonstrating both Mott -- Hubbard
and Anderson metal -- insulator transitions in two -- dimensional systems of the
finite size, allowing us to construct the complete zero -- temperature phase
diagram of paramagnetic Anderson -- Hubbard model. Localization length in our
approximation is practically independent of the strength of Hubbard correlations.
However, the divergence of localization length in finite size 
two -- dimensional system at small disorder signifies the existence of an
effective Anderson transition. 

\end{abstract}


\maketitle

\newpage

\section{Introduction}

The study of disordered electronic systems with the account of interaction
effects belongs to the central problems of the modern condensed matter theory
\cite{Lee85}. According to the scaling theory of localization \cite{AALR79} 
there is no metallic state in two -- dimensional (2D) systems, all electronic 
states are localized at the smallest possible disorder. This prediction was made 
first for noninteracting 2D systems, soon after it was shown that the weak
electron -- electron interaction in most cases enhance localization \cite{AAL80}. 
Experiments performed in early 80's on different 2D systems \cite{2Dexp}
essentially confirmed these predictions. However, some theoretical works
produced an evidence of rather different possibility \cite{ma} for 2D systems to
evolve to the state with even infinite metallic -- like conductivity at zero
temperature ($T=0$) in case of weak disorder and sufficiently strong correlations.
Experimental observation of metal -- insulator transition (MIT) in 2D systems
with weak enough disorder but strong correlations (low electronic densities)
\cite{Krav04}, which apparently contradicted the predictions of the scaling
theory of localization, stimulated extensive theoretical studies with no
widely accepted solution up to now (see the review in Ref.~\onlinecite{AKS01}).

One of the basic models allowing for the simultaneous account of both
strong enough electronic correlations, leading to Mott MIT
transition \cite{Mott90}, and effects of strong disorder, leading to Anderson
MIT \cite{Anderson58}, is the Anderson -- Hubbard model, intensively studied in 
recent years \cite{Dobrosavljevic97, Dobrosavljevic03, BV, HubDis, ShapiroPRB08, Shapiro08, PB09}.

In Refs. \cite{Dobrosavljevic97, Dobrosavljevic03, BV} three -- dimensional
(3D) Anderson -- Hubbard model was analyzed with dynamical mean field theory
(DMFT), which is extensively used in the theory of strongly correlated
electrons \cite{MetzVoll89,vollha93,pruschke,georges96}. However, disorder
effects were mostly taken into account via the average density of states and
the coherent potential approximation (CPA) \cite{ulmke95,vlaming92}, 
which misses the effects of Anderson localization. To overcome this problem
Dobrosavljevi\'c and Kotliar \cite{Dobrosavljevic97} have proposed the DMFT 
approach, where the solution of self -- consistent stochastic DMFT equations
were used to calculate geometrically averaged local density of states.
This approach was further developed in Refs. \cite{Dobrosavljevic03, BV} with
DMFT account for Hubbard correlations, which leads to highly nontrivial
phase diagram of 3D paramagnetic Anderson -- Hubbard model \cite{BV},
containing the phases of correlated metal, Mott insulator and correlated
Anderson insulator. However, the major problem of the approach of Refs.
\cite{Dobrosavljevic97, Dobrosavljevic03, BV} is its practical inability of
direct calculations of conductivity, which actually determines MIT itself.

In the previous work \cite{HubDis} we have studied the 3D paramagnetic
Anderson -- Hubbard model using our recently developed DMFT+$\Sigma$ 
approximation \cite{JTL05, PRB05, FNT06, PRB07}, which conserves the standard
single impurity DMFT approach, taking into account the local Hubbard
correlations, allowing the use the usual `` impurity solvers'' like NRG 
\cite{NRG, BPH, Bull}, at the same time allowing to include additional
(local or nonlocal) interactions. Strong disorder was accounted via some
generalization of the self -- consistent theory of localization
\cite{VW, WV, MS, MS86, VW92, Diagr}. In the framework of this approach we
have been able not only to reproduce the phase diagram qualitatively similar to
that obtained in Ref.~\onlinecite{BV}, but also calculate the dynamic (optical)
conductivity for the wide frequency range.

In Ref.~\onlinecite{Shapiro08} the Hubbard -- Anderson model was studied both for
3D and 2D systems. As the main mechanism leading to delocalization a kind of
``screening'' of the random (disorder) potential by local Hubbard interaction
was introduced \cite{ShapiroPRB08}. Then the Anderson -- Hubbard model was
reduced to an effective single -- particle Anderson model with renormalized
distribution of local site energies, which was calculated in the atomic limit.
All the other effects of electronic correlations were neglected.
Strong disorder effects were accounted within self -- consistent theory of
localization. In this approach the authors obtained the significant growth of
localization length with growing Hubbard interaction in 2D. However, localization
length itself remained finite, the system being localized at smallest possible
disorder, so that Anderson in 2D is still absent. Similar result was obtained
also in numerical simulations of 2D Anderson -- Hubbard model in 
Ref.~\onlinecite{PB09}.

In this work we present a direct generalization of the method used in Ref.~\onlinecite{HubDis}
to the case of 2D systems. We use the DMFT+$\Sigma$ approach to
calculate DOS, optical conductivity, localization length and construct the
phase diagram of 2D paramagnetic Anderson -- Hubbard model with strong
electronic correlations and strong disorder. Strong correlations are taken
into account via DMFT, while disorder effects are treated by the appropriate
generalization of the self -- consistent theory localization.

The paper is organized as follows: in section \ref{leng_intro} we present a brief
description of our DMFT+$\Sigma$ approximation as applied to disordered 
Hubbard model. In section \ref{opt_cond} we formulate the basic 
DMFT+$\Sigma$ expressions for optical conductivity and self -- consistency
equation for the generalizes diffusion coefficient. Our results for DOS,
optical conductivity and localization length are given in section 
\ref{results}, where we also analyze the phase diagram of 2D disordered
Hubbard model and briefly discuss the optical sum rule within our approach.
Finally we present a short conclusion, which includes the discussion of problems
yet to be solved.

\section{Basics of DMFT+$\Sigma$ approach.}
\label{leng_intro}

In the following we consider paramagnetic disordered Anderson -- Hubbard
model at half -- filling for arbitrary correlations and disorder. This model
treats both Mott -- Hubbard and Anderson MIT on the same footing. 
The Hamiltonian of the model can be written as:
\begin{equation}
H=-t\sum_{\langle ij\rangle \sigma }a_{i\sigma }^{\dagger }a_{j\sigma
}+\sum_{i\sigma }\epsilon _{i}n_{i\sigma }+U\sum_{i}n_{i\uparrow
}n_{i\downarrow },  
\label{And_Hubb}
\end{equation}
where $t>0$ is nearest neighbor transfer integral, $U$ is the local Hubbard
repulsion, $n_{i\sigma }=a_{i\sigma }^{\dagger }a_{i\sigma }$ 
is electron number operator at a given site $i$, 
$a_{i\sigma }$ ($a_{i\sigma }^{\dagger}$) is annihilation (creation) operator
for an electron with spin $\sigma$, local energies $\epsilon _{i}$ are assumed
to be randomly and independently distributed on different lattice sites. 
To simplify diagram technique in the following we assume $\epsilon _{i}$ 
distribution to be Gaussian:
\begin{equation}
\mathcal{P}(\epsilon _{i})=\frac{1}{\sqrt{2\pi}\Delta}\exp\left(
-\frac{\epsilon_{i}^2}{2\Delta^2}
\right)
\label{Gauss}
\end{equation}
Here $\Delta$ serves as disorder parameter and Gaussian random field 
(``white'' noise) of energy levels $\epsilon _{i}$ at different lattice sites
induces ``impurity'' -- like scattering, leading to the standard diagram
technique for calculations of the averaged Green's functions \cite{Diagr}.

DMDF+$\Sigma$ approach, initially proposed in Refs. \cite{JTL05,PRB05,FNT06,PRB07}
as a simple method to include nonlocal interactions (fluctuations) into
the standard (local) DMFT scheme, is also very convenient for the account 
in DMFT of any additional interaction (local or nonlocal) of arbitrary nature.

In DMFT+$\Sigma$ approximation we choose the single -- particle Green's
function in the following form:
\begin{equation}
G_{\bf p}(\varepsilon)=\frac{1}{\varepsilon+\mu-\varepsilon({\bf p})-\Sigma(\varepsilon)
-\Sigma_{\bf p}(\varepsilon)},
\label{Gk}
\end{equation}
where $\varepsilon({\bf p})$ is the ``bare'' electron spectrum, 
$\Sigma(\varepsilon)$ is the local (DMFT) self -- energy due to Hubbard
interactions, while $\Sigma_{\bf p}(\varepsilon)$ is an ``external'' 
(in general case momentum dependent) self -- energy due to some other 
interaction. The main assumption of our approach (both its advantage and
deficiency) is precisely in this additive form (neglect of interference effects) 
for the total self -- energy in (\ref{Gk}) \cite{JTL05,PRB05,FNT06,PRB07}, 
which allows us to conserve the standard form of self -- consistent DMFT 
equations \cite{georges96} with two major generalizations. First of all, at
each iteration of DMFT -- loop we recalculate an ``external'' self -- energy 
$\Sigma_{\bf p}(\mu,\varepsilon,[G_{\bf p}(\varepsilon)])$ within some
(approximate) scheme, taking into account the ``external'' interaction (in the
present case that due to disorder scattering). Secondly, the local Green's
function for an effective DMFT -- impurity problem is defined as:
\begin{equation}
G_{ii}(\varepsilon)=\frac{1}{N}\sum_{\bf p}\frac{1}{\varepsilon+\mu
-\varepsilon({\bf p})-\Sigma(\varepsilon)-\Sigma_{\bf p}(\varepsilon)},
\label{Gloc}
\end{equation}
at each step of the standard DMFT procedure.
Finally we obtain the desired Green's function in the form of Eq. (\ref{Gk}), 
where  $\Sigma(\varepsilon)$ and $\Sigma_{\bf p}(\varepsilon)$ are self --
energies obtained at the end of our iteration procedure.

For $\Sigma_{\bf p}(\varepsilon)$ appearing due to disorder scattering we shall
use the simple one -- loop contribution, neglecting diagrams with ``crossing''
interaction lines, i.e. self -- consistent Born approximation 
\cite{Diagr}, which in the case of Gaussian disorder (\ref{Gauss}) leads to
the usual expression:
\begin{equation}
\Sigma_{\bf p}(\varepsilon)=\Delta^2\sum_{\bf p}G(\varepsilon,{\bf p})
\equiv \Sigma_{imp}(\varepsilon)
\label{BornSigma}
\end{equation}
and ``external'' self -- energy in this case is ${\bf p}$-independent (local).

\section{Optical conductivity in DMFT+$\Sigma$ approach.}
\label{opt_cond}

It is obvious that calculations of optical (dynamic) conductivity provide the
direct way to study MIT as frequency dependence of conductivity, as well as
its static value at zero frequency of an external field, allows the clear
distinction between metallic and insulating phases (at temperature $T=0$ ).

\begin{figure}[t]
\includegraphics[clip=true,width=0.6\columnwidth]{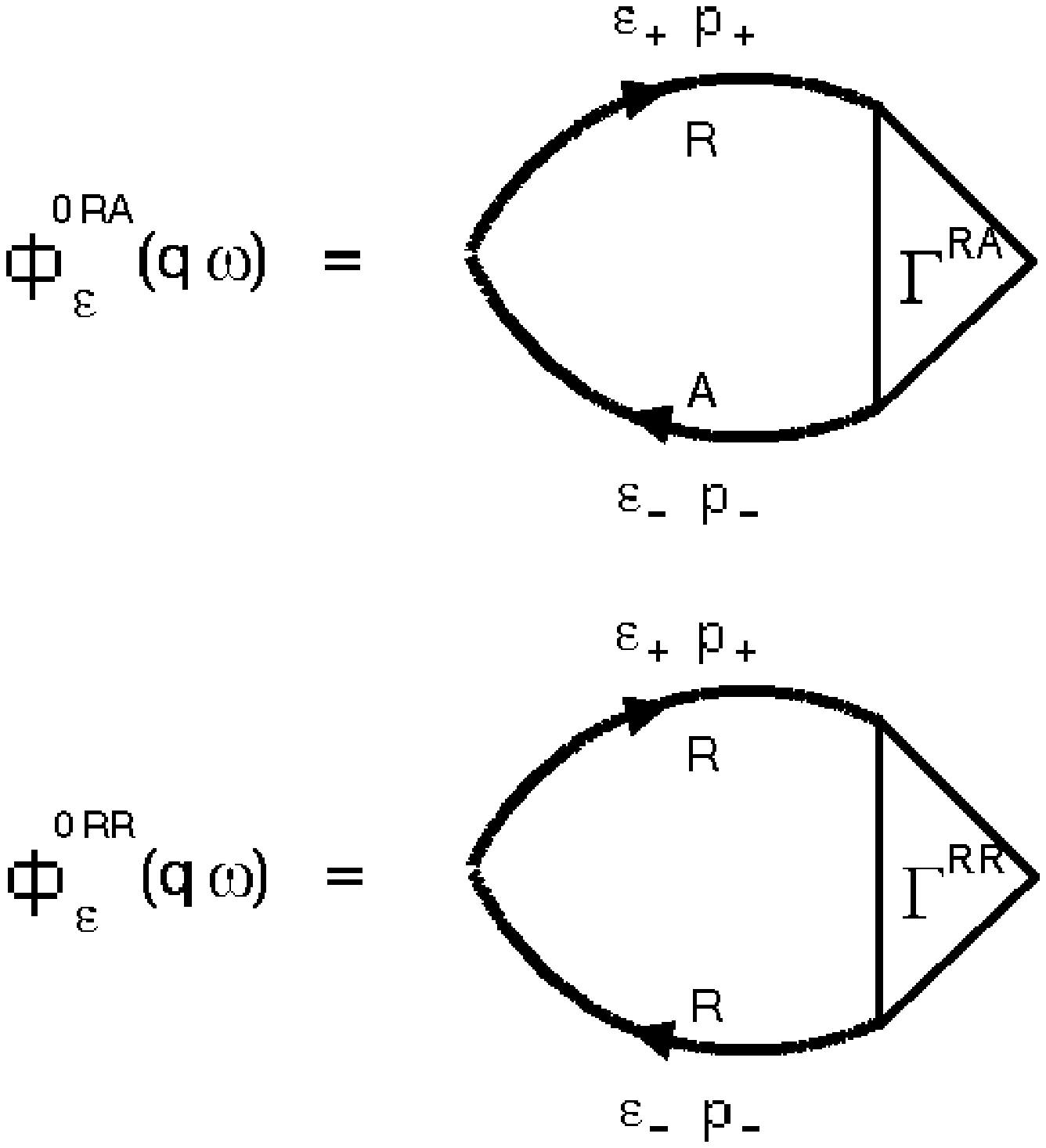}
\caption{Diagrammatic representation of  
$\Phi^{0RA}_{\varepsilon}(\omega,{\bf q})$ è  $\Phi^{0RR}_{\varepsilon}
(\omega,{\bf q})$.} 
\label{loop_fi} 
\end{figure}

Local nature of irreducible self -- energy in  DMFT allows to reduce the
problem of calculation of optical conductivity to calculation of the usual
particle -- hole loop without DMFT vertex corrections due to local Hubbard
interaction \cite{PRB07,HubDis}. The final expression for the real part of
optical conductivity obtained in this way in Refs. \cite{PRB07,HubDis}, 
takes the following form:
\begin{eqnarray}
{\rm{Re}}\sigma(\omega)=\frac{e^2\omega}{2\pi}
\int_{-\infty}^{\infty}d\varepsilon\left[f(\varepsilon_-)
-f(\varepsilon_+)\right]\times\nonumber\\
\times{\rm{Re}}\left\{\phi^{0RA}_{\varepsilon}(\omega)\left[1-
\frac{\Sigma^R(\varepsilon_+)-\Sigma^A(\varepsilon_-)}{\omega}\right]^2-
\right.\nonumber\\
\left.-\phi^{0RR}_{\varepsilon}(\omega)\left[1-
\frac{\Sigma^R(\varepsilon_+)-\Sigma^R(\varepsilon_-)}{\omega}\right]^2
\right\},
\label{cond_final}
\end{eqnarray}
Here $f(\varepsilon)$ is Fermi distribution,
$\varepsilon_{\pm}=\varepsilon\pm\frac{\omega}{2}$ and
\begin{equation}
\phi^{0RR(RA)}_{\varepsilon}(\omega)=\lim_{q\to 0}
\frac{\Phi^{0RR(RA)}_{\varepsilon}(\omega,{\bf q})-
\Phi^{0RR(RA)}_{\varepsilon}(\omega,0)}{q^2},
\label{fi0RA_func}
\end{equation}
where we have introduced two -- particle loops:
\begin{eqnarray}
\Phi^{0RR(RA)}_{\varepsilon}(\omega,{\bf q})=\sum_{\bf p}
G^R(\varepsilon_+,{\bf p_+})G^{R(A)}(\varepsilon_-,{\bf p_-})\nonumber\\
\Gamma^{RR(RA)}(\varepsilon_-,{\bf p}_-;\varepsilon_+,{\bf p}_+),
\label{PhiRA}
\end{eqnarray}
represented diagrammatically in Fig. \ref{loop_fi} with
${\bf p}_{\pm}={\bf p}\pm\frac{\bf q}{2}$ and $R$ and $A$ indices corresponding 
to retarded and advanced Green's functions. The vertices
$\Gamma^{RR(RA)}(\varepsilon_-,{\bf p}_-;\varepsilon_+,{\bf p}_+)$ 
contain all vertex corrections due to disorder scattering, but do not
include vertex corrections due to Hubbard interaction.

Thus, the problem is much simplified. To calculate optical conductivity in
DMFT+$\Sigma$ approximation we have only to solve single -- particle problem
to determine the local self -- energy $\Sigma(\varepsilon_{\pm})$ with the help
of DMFT+$\Sigma$ procedure, described above, while the nontrivial contribution
of disorder scattering enters via $\phi^{0RR(RA)}$ of Eq. (\ref{fi0RA_func}), 
which may be calculated in some appropriate approximation.
In fact,  $\phi^{0RR(RA)}$ contain only disorder scattering, though using
as the ``bare'' the Green's functions including the DMFT self -- energies,
already determined with the help of DMFT+$\Sigma$ procedure. 
Eq. (\ref{cond_final}) guarantees the effective interpolation between the case 
of strong correlations in the absence of disorder and the case of pure disorder
in the absence of Hubbard correlations.

The most important $\Phi^{0RA}_{\varepsilon}(\omega,{\bf q})$ loop may be
calculated  using the basic approach of the self -- consistent theory of
localization \cite{VW,WV,MS,MS86,VW92,Diagr} with some generalizations
accounting for Hubbard interaction within the DMFT+$\Sigma$ approach \cite{HubDis}.

The rest is the direct generalization of the scheme proposed in Ref.~\onlinecite{HubDis}
for the two -- dimensional case. Here we present only some basic points of the
approach of Ref.~\onlinecite{HubDis}, stressing important differences due to
two -- dimensionality of the model.

In RA-channel the two -- particle loop 
$\Phi^{0RA}_{\varepsilon}({\bf q},\tilde\omega)$ possesses a diffusion -- like
contribution:
\begin{equation}
\Phi^{0RA}_{\varepsilon}({\bf q},\tilde\omega)=
\frac{-\sum_{\bf p}\Delta G_{\bf p}}
{\tilde\omega+iD(\omega)q^2}, 
\label{FiRAfull} 
\end{equation}
where $\Delta G_{\bf p}=G^R(\varepsilon_+,{\bf p})-G^A(\varepsilon_-,{\bf p})$. 
The important difference with noninteracting case is contained in
\begin{eqnarray}
&&\tilde\omega=\varepsilon_+-\varepsilon_-
-\Sigma^R(\varepsilon_+)+\Sigma^A(\varepsilon_-)= \nonumber\\
&&\omega-\Sigma^R(\varepsilon_+)+\Sigma^A(\varepsilon_-)
\equiv
\omega-\Delta\Sigma^{RA}(\omega)
\label{tomega}
\end{eqnarray}
which replaces the usual $\omega$-term in the denominator of the standard
expression for $\Phi^{0RA}_{\varepsilon}(\omega,{\bf q})$ \cite{Diagr}.

Then (\ref{cond_final}) can be rewritten as:
\begin{eqnarray}
{\rm{Re}}\sigma(\omega)=\frac{e^2\omega}{2\pi}
\int_{-\infty}^{\infty}d\varepsilon\left[f(\varepsilon_-)
-f(\varepsilon_+)\right]\times\nonumber\\
\times{\rm Re}\left\{\frac{i\sum_{\bf p}\Delta G_{\bf p}D(\omega)}
{\omega^2} - 
\phi^{0RR}_{\varepsilon}(\omega)\left[1-
\frac{\Delta\Sigma^{RR}(\omega)}{\omega}\right]^2\right\}\nonumber\\
\label{con_fin}
\end{eqnarray}
The second term in (\ref{con_fin}), which is actually irrelevant at small
$\omega$, can be obtained from (\ref{fi0RA_func}) calculating
$\Phi^{0RR}_{\varepsilon}(\omega,{\bf q})$ in the usual ``ladder'' approximation.

Repeating the derivation scheme of the self -- consistent theory of localization
presented in Ref.~\onlinecite{HubDis}, we obtain the following equation for the
generalized diffusion coefficient:
\begin{eqnarray}
D(\omega)=i\frac{<v>^2}{d}\left\{\tilde\omega-
\Delta\Sigma_{imp}^{RA}(\omega)\right.+\nonumber\\
\left.+\Delta^4\sum_{\bf p}(\Delta G_{\bf p})^2\sum_{\bf q}
\frac{1}{\tilde\omega+iD(\omega)q^2}\right\}^{-1},
\label{Dsc}
\end{eqnarray}
where $d=2$ is spatial dimension and
$\Delta\Sigma^{RA}_{imp}(\omega)=\Sigma_{imp}^R(\varepsilon_+)
-\Sigma_{imp}^A(\varepsilon_-)$ is determined by disorder scattering.
The average velocity $<v>$, well approximated by Fermi velocity, is defined as:
\begin{equation}
<v>=\frac{\sum_{\bf p}|{\bf v}_{\bf p}|\Delta G_{\bf p}}
{\sum_{\bf p}\Delta G_{\bf p}};\\\ {\bf v}_{\bf p}=\frac{\partial\epsilon({\bf p})}
{\partial{\bf p}},
\label{veloc1}
\end{equation}
Due to the limits of diffusion approximation summation over $q$ in (\ref{Dsc}) 
should be limited by \cite{MS86,Diagr}:
\begin{equation}
q<k_0=Min \{l^{-1},p_F\}
\label{cutoffu}
\end{equation}
where $l=<v>/2\gamma$ is the mean -- free path due to elastic scattering
($\gamma$ is the scattering rate due to disorder), $p_F$ is Fermi momentum. 
In our two -- dimensional model Anderson localization takes place at
infinetisimal disorder. However, for small disorder localization length is
exponentially large, so that the size of the sample becomes important.
The sample size $L$ may be introduced into the self -- consistent theory of
localization as a cutoff of diffusion pole contribution at small $q$
\cite{VW, WV}, i.e. for:
\begin{equation}
q\sim k_L=1/L.
\label{cutoffd}
\end{equation}

Eq. (\ref{Dsc}) for the generalized diffusion coefficient reduces to a
transcendental equation, which is easily solved by iterations for each value of
$\tilde\omega$, taking into account that for $d=2$ and cutoffs defined by
Eqs. (\ref{cutoffu}), (\ref{cutoffd}) the sum over $q$ in (\ref{Dsc}) 
reduces to:
\begin{equation}
\sum_{\bf q}\frac{1}{\tilde\omega+iD(\omega)q^2}=
\frac{1}{i2\pi D(\omega)}
\int_{\frac{k_L}{k_0}}^{1}\frac{ydy}{y^2+\frac{\tilde\omega}{iD(\omega)k_0^2}}=
\label{Sumq}
\end{equation}
\begin{equation}
=\frac{1}{i4\pi D(\omega)}
\ln
{\left(
\frac{1-\frac{i\tilde\omega}{D(\omega)k_0^2}}
{(\frac{k_L}{k_0})^2-\frac{i\tilde\omega}{D(\omega)k_0^2}}
\right)}
\\ \nonumber
\end{equation}

Solving Eq. (\ref{Dsc}) for different values of our model parameters and using
Eq. (\ref{con_fin}) we can directly calculate the optical (dynamic) conductivity 
in different phases of the Anderson -- Hubbard model.

For $\omega \rightarrow 0$ (and at the Fermi level ($\varepsilon=0$) 
obviously also $\tilde\omega \rightarrow 0$), in Anderson insulator phase we
obtain the localization behavior of the generalized diffusion coefficient
\cite{VW, WV,Diagr}:
\begin{equation}
D(\omega)=-i\tilde\omega {R_{loc}}^2.
\label{DR}
\end{equation}
After the substitution of (\ref{DR}) into (\ref{Dsc}) we get an equation,
determining the localization length $R_{loc}$:
\begin{equation}
{R_{loc}}^2=-\frac{<v>^2}{d\Delta^4}
\left\{\sum_{\bf p}(\Delta G_{\bf p})^2\sum_{\bf q}
\frac{1}{1+{R_{loc}}^2 q^2}\right\}^{-1}.
\label{Rl2}
\end{equation}
where the sum over $q$ is defined by (\ref{Sumq}). As we shall see in the
following, for an infinite two -- dimensional system ($L\to\infty$) localization
length, determined by Eq. (\ref{Rl2}) remains finite (though exponentially
large) for the smallest possible disorder, signifying the absence of Anderson
transition. However, for the finite size systems localization length diverges
at some critical disorder, determined for each value of the system size $L$.
Qualitatively, this critical disorder is determined from the condition of
localization length (in infinite system) becoming of the order of characteristic
sample size $R_{loc}^{L\to\infty}\sim L$. Thus, in finite two -- dimensional
systems the Anderson transition and metallic phase do exist for disorder
strength lower, than this critical disorder. In the following, this kind of
metallic phase will be referred to as a phase of the ``correlated metal'' in
finite 2D systems.

\section{Main results}
\label{results}

Below we present the results of extensive numerical calculations for 2D
Anderson -- Hubbard model on the square lattice with rectangular ``bare''
DOS, corresponding to the bandwidth $W=2D$:
\begin{equation}
N_0(\varepsilon)=
\left\{
\begin{array}{ll}
\frac{1}{2D} & \quad |\varepsilon|\leq D\\ 
0 & \quad |\varepsilon| > D
\end{array}.
\right.
\label{DOS}
\end{equation}
The choice of this model DOS is dictated by its 2D nature. 

Everywhere below we give the values of DOS in units of the number of states
per energy interval, per lattice cell of the volume $a^2$ ($a$ -- lattice
parameter), per one spin projection.

As we concentrate on half -- filled case, Fermi level is always assumed to be
at zero energy.

As ``impurity solver'' for DMFT effective impurity problem we have used the
reliable numerical renormalization group (NRG) approach \cite{NRG,BPH,Bull}. 
Calculations were made for low enough temperature $T\sim 0.001D$, so that
temperature effects in DOS and conductivity are just negligible.

Now we present only most typical results.

\subsection{Evolution of the density of states}

\begin{figure}[t]
\includegraphics[clip=true,width=1.0\columnwidth]{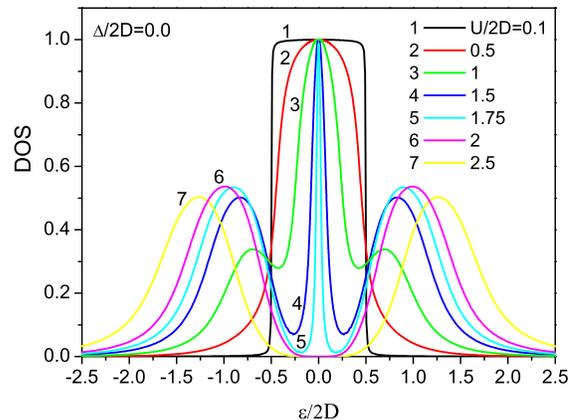}
\caption{Density of states of the half -- filled Anderson -- Hubbard model for
different values of $U$ and $\Delta=0$ (no disorder).} 
\label{DOSU} 
\end{figure} 

In Fig. \ref{DOSU} we show evolution of the DOS with the growth of Hubbard
interaction $U$ in the absence of disorder. At small $U$ (curve 1  in Fig.\ref{DOSU}) èìååì 
we observe practically rectangular DOS almost coinciding with the ``bare'' one.
As $U$ grows a typical three peak structure of DOS appears 
\cite{pruschke,georges96,Bull} (curves 3,4,5 in Fig.\ref{DOSU}):  a narrow
quasiparticle peak at the Fermi level with upper and lower Hubbard bands at
$\varepsilon\sim\pm U/2$. Quasiparticle peak narrows as $U$ grows in metallic
phase, disappearing at Mott MIT for $U=U_{c2}\approx1.83W$. With further growth
of $U$ (curves 6,7 in Fig.\ref{DOSU}) dielectric gap opens at the Fermi level.

\begin{figure}[t]
\includegraphics[clip=true,width=1.0\columnwidth]{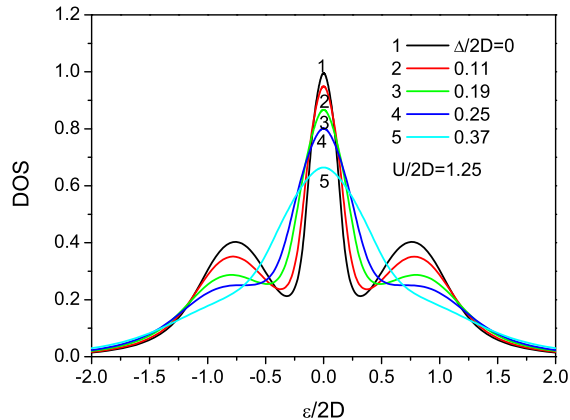}
\caption{Density of states of the half -- filled Anderson -- Hubbard model
for different values of disorder $\Delta$ and $U/2D=1.25$, typical for 
correlated metal (in the absence of disorder).} 
\label{metDOS} 
\end{figure} 

In Fig. \ref{metDOS} we show the results for DOS obtained at relatively
weak correlation strength $U=1.25W$ ($W=2D$), so that the system is rather far
fron the Mott transition, but for the wide range of disorder strength $\Delta$. 
We observe typical widening of the band with appropriate suppression of DOS as
disorder grows.

\begin{figure}[t]
\includegraphics[clip=true,width=1.0\columnwidth]{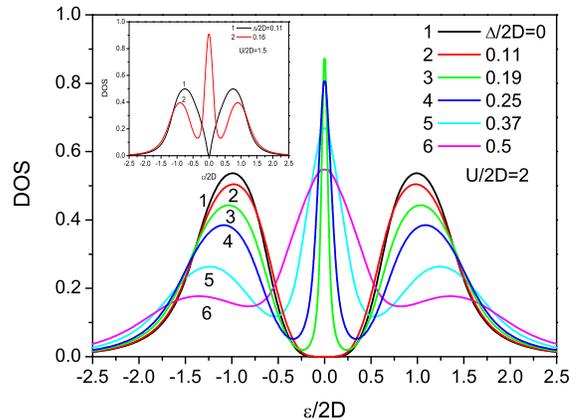}
\caption{Density of states of the half -- filled Anderson -- Hubbard model
for different values of disorder $\Delta$ and $U/2D=2$, typical
for Mott insulator (in the absence of disorder). 
At the insert -- restoration of quasiparticle band 
by disorder in coexistence (hysteresis) region for $U=1.5D$, 
obtained from Mott insulator with decreasing $U$.} 
\label{insDOS} 
\end{figure}

In Fig. \ref{insDOS} DOS evolution is shown as disorder $\Delta$ grows at
$U=2W$, typical for Mott insulator in the absence of disorder. It can be seen
that the growth of disorder leads to restoration of the quasiparticle peak
in DOS. Similar unusual behavior of DOS (closing of dielectric gap by disorder)
was first noted in 3D systems \cite{HubDis}. However, in the present 2D case
it does not, in general, signify the transition to the correlated metal phase,
at least for the infinite systems we are, in fact, dealing with correlated
Anderson insulator (cf. below).

The physical reason for this unusual restoration of the quasiparticle peak in
DOS is clear. Controlling parameter for appearance or disappearance of 
quasiparticle peak in DMFT in the absence of disorder is the ratio of
Hubbard interaction $U$ and the ``bare'' bandwidth $W=2D$. Disordering leads to
the growth of the effective bandwidth $W_{eff}$ (in the absence of Hubbard
interaction) and appropriate suppression of $U/W_{eff}$ ratio, which obviously
leads to the restoration of quasiparticle band in our model. In more details
this qualitative picture will be discussed in Section \ref{phd}, where our
results for DOS will be used in construction of the phase diagram of 2D
Anderson -- Hubbard model.

It is well known, that hysteresis behavior of DOS is obtained for 
Mott--Hubbard transition if we perform DMFT calculations with $U$ decreasing
from insulating phase \cite{georges96,Bull}. Mott insulator phase survives 
for the values of $U$ well inside the correlated metal phase, obtained with
the increase of $U$. Metallic phase is restored at $U_{c1}\approx 1.42W$. The
values of $U$ from the interval $U_{c1}<U<U_{c2}$ are usually considered 
as belonging to coexistence region of metallic and (Mott) insulating phases,
with metallic phase being thermodynamically more stable 
\cite{georges96,Bull}. In the coexistence region disorder increase also leads to the
restoration of quasiparticle peak in the DOS  (see insert of Fig.\ref{insDOS}).

\subsection{Optical conductivity: Mott -- Hubbard and Anderson transitions.}
\label{cond}

The real part of optical conductivity was calculated for different combinations
of parameters of the model, directly from Eqs. (\ref{con_fin}) and (\ref{Dsc}), 
using the results of DMFT+$\Sigma$ procedure for single -- particle 
characteristics. The values of conductivity below are given in natural units of
${e^2}/{\hbar }$.

In the absence of disorder we just reproduce the results of the standard DMFT
with optical conductivity characterized by the usual Drude peak at zero
frequency and a wide maximum at $\omega\sim U$, corresponding to transitions
to the upper Hubbard band. As $U$ grows Drude peak is suppressed and
disappear at Mott MIT, when only remaining contribution is due to
transitions through the Mott -- Hubbard gap. 

Introduction of disorder leads to qualitative change of frequency behavior of
conductivity. Below we mainly present results obtained for the same values of
$U$ and $\Delta$, which were used above to illustrate evolution of DOS.

\begin{figure}[t]
\includegraphics[clip=true,width=1.0\columnwidth]{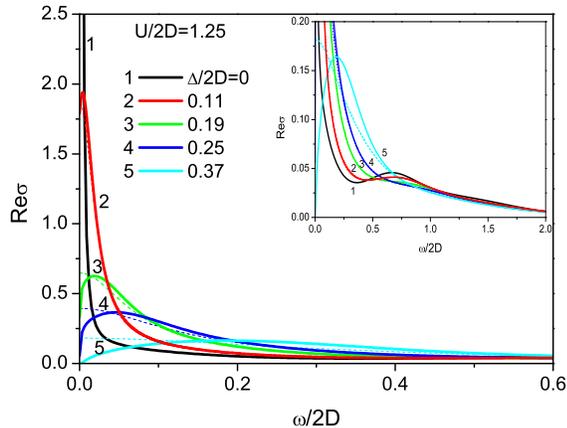}
\caption{Real part of dynamic conductivity of the half -- filled Anderson -- Hubbard
model for different values of disorder $\Delta$ and $U/2D=1.25$, typical for
correlated metal (in the absence of disorder). At the insert -- the same, but
in wider frequency range. Thin dashed lines represent the results of ladder
approximation.} 
\label{met_cond} 
\end{figure} 

In Fig. \ref{met_cond} we show the real part of optical conductivity
in 2D half -- filled Anderson -- Hubbard model for different disorder strengths
$\Delta$ and $U=1.25W$, when the system is far from Mott MIT. Thin dashed lines
in Fig. \ref{met_cond} (as well as in the following figures) we show the results
of the ``ladder'' approximation. In 2D model under consideration conductivity
at zero frequency is always zero, and in contrast to 3D case \cite{HubDis},
even for the weakest possible disorder the peak in optical conductivity is at
finite frequency. In the ``ladder'' approximation, which does not contain
localization corrections, conductivity at $\omega=0$ is finite. Optical
transitions to the upper Hubbard band at $\omega\sim U$ are practically
unobservable in these data, only at the insert in Fig. \ref{met_cond}, 
where we show the data for the wide frequency range, we can observe some
weak maximum on curves 1 and 2, corresponding to transitions to the upper
Hubbard band. 

\begin{figure}[t]
\includegraphics[clip=true,width=1.0\columnwidth]{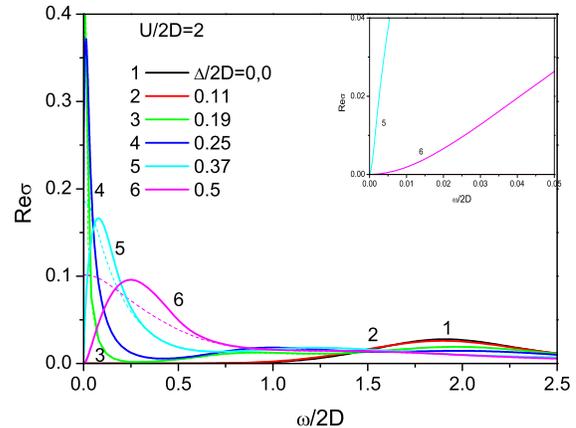}
\caption{Real part of dynamic conductivity of the half -- filled
Anderson -- Hubbard model for different values of disorder
$\Delta$ and $U/2D=2$, typical for Mott insulator (in the absence of disorder).
Curves 1,2 correspond to Mott insulator, while curves
3-6 correspond to correlated Anderson insulator. At the insert -- localization
behavior of conductivity. Thin dashed lines -- ladder approximation.} 
\label{ins_cond} 
\end{figure} 

In Fig. \ref{ins_cond} we present the real part of optical conductivity
for different disorder strengths $\Delta$ and $U=2W$, typical for Mott
insulator. It can be seen from Fig. \ref{ins_cond} that for small disorder
we are in Mott insulator phase (curves 1,2), and with the growth of disorder
in the absence of Anderson localization (cf. thin lines corresponding to 
``ladder'' approximation) we would be entering the metallic phase. 
However, in our model localization takes place at infinetesimal disorder and
we are actually entering Anderson insulator phase, with conductivity going to
zero at zero frequency. Data in the frequency range corresponding to localization
behavior of conductivity $\sigma(\omega)\sim{\omega}^2$ are shown at the insert
in Fig. \ref{ins_cond} for curve 5 and 6, corresponding to large enough disorder.
At small disorders the frequency region with localization behavior of 
conductivity is exponentially small\footnote{This region corresponds to
frequencies \cite{VW,WV} $\omega\ll\omega_c\sim\frac{D^3}{\Delta^2}/(\frac{R_{loc}}{a})^2$.}
(which is due to the exponential growth of localization length at small disorder,
cf. Fig.\ref{RsRloc}) and is practically unobservable.

\begin{figure}[t]
\includegraphics[clip=true,width=1.0\columnwidth]{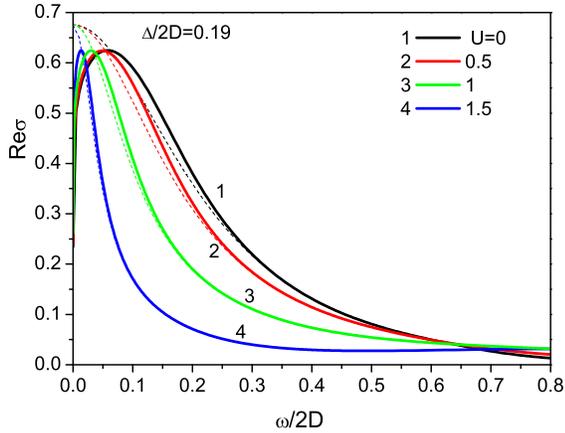}
\caption{Real part of dynamic conductivity of the half -- filled 
Anderson -- Hubbard model for different values of $U$ and $\Delta/2D=0.19$.
Thin dashed lines -- ladder approximation.} 
\label{u_cond} 
\end{figure} 

Dependence of optical conductivity on $U$ is illustrated in Fig. \ref{u_cond}.
The growth of $U$ shifts localization peak in conductivity to lower frequencies
and leads to its narrowing. Apparently this is related to the appropriate
suppression of quasiparticle peak width in DOS. The value of conductivity at
the maximum is $U$ independent. It is interesting to note that for frequencies 
larger than the maximum position the growth of $U$ suppresses conductivity, 
while for the frequencies lower than the maximum position the growth of $U$ 
enhances conductivity (playing in a sense against localization).

To confirm self -- consistency of our approach to conductivity calculations
we conclude this section with discussion of the optical sum rule, which relates
single -- particle and two -- particle characteristics  \cite{sr08}.

The single -- band Kubo sum rule \cite{Kubo} for dynamic conductivity can be
written as:
\begin{equation}
W_{opt} = \int_0^{\infty} Re \sigma(\omega) d\omega =
 \frac{\pi e^2}{2}\sum_{\bf p}
\frac{\partial^2\varepsilon_p}{\partial p_x^2}n_p, 
\label{1}
\end{equation}
where $n_p$ is single -- particle momentum distribution function, determined by  
{\em interacting} retarded electron Green's function $G^R(\varepsilon,{\bf  p})$:  
\begin{equation} 
n_p=-\frac{1}{\pi}\int_{-\infty}^{\infty} d\varepsilon 
f(\varepsilon)Im G^R(\varepsilon,{\bf p})
\label{dist_fun}
\end{equation}
where $f(\varepsilon)$ is Fermi distribution.

In Table I we show calculated values of the r.h.s. and of the l.h.s of 
Eq. (\ref{1}) for $U=1.5W$. It is clearly seen that the optical (\ref{1}) sum 
rule is fulfilled within our numerical accuracy.

\begin{figure}[t]
\includegraphics[clip=true,width=1.0\columnwidth]{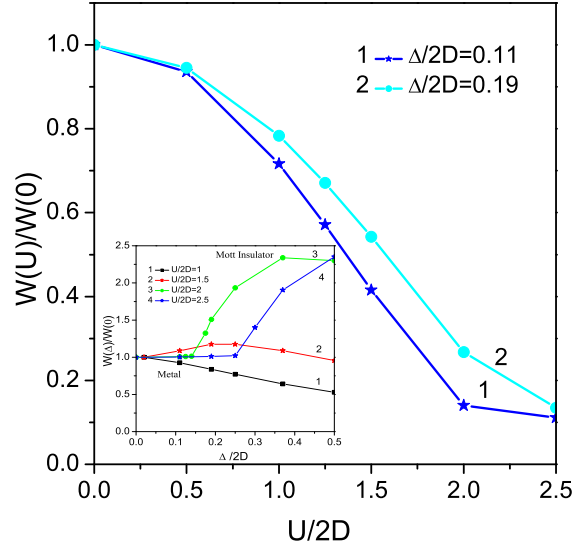}
\caption{Dependence of the normalized optical integral of Anderson -- Hubbard
model on $U$ for different values of disorder $\Delta$. At the insert --
similar dependence on $\Delta$ for different values of $U$.
Curves 1,2 -- ``correlated metal'', transforming into Anderson insulator.
Curves 3,4 -- Mott insulator, obtained with the growth of $U$ from
``correlated metal'' or Anderson insulator.} 
\label{W_opt} 
\end{figure}

\begin{table}
\label{Tab1}
\caption {Check of the single band optical sum rule in Anderson -- Hubbard
model. Optical integral is given in units of $\frac{2e^2}{\hbar}D$.}  
\begin{tabular}{| c | c | c |}
\hline
$\Delta/2D$    &  
$\frac{\pi e^2}{2}\sum_{\bf p}\frac{\partial^2\varepsilon_p}{\partial p_x^2}n_p$  & $W_{opt} = \int_0^{\infty} Re \sigma(\omega) d\omega $ \\ 
\hline 
$0.19$  & 0.099 & 0.098 \\ 
\hline 
$0.25$  & 0.099 & 0.098 \\ 
\hline
$0.37$ & 0.092 & 0.091  \\
\hline
$0.5$ & 0.081  & 0.082 \\
\hline
\end{tabular}
\end{table}
Very often the optical sum rule is understood as the equality of the optical
integral $W_{opt}$ to the ``universal'' value of $\frac{\omega_{pl}^2}{8}$, where
$\omega_{pl}$ is plasma frequency, which is strictly speaking is not correct in 
single band case and in this sense we may speak of optical sum rule ``violation''.
In fact the optical integral depends on parameters of the model, e.g. on
Hubbard interaction and disorder (Fig. \ref{W_opt}). The growth of $U$ 
significantly suppresses the value of optical integral. Dependence on disorder
strength  $\Delta$  is also important, in particular disorder induced
transition from Mott to Anderson insulator we observe a kind of discontinuity of
optical integral (curves 3,4 at the insert in Fig. \ref{W_opt}).

\subsection{Localization length and phase diagram of 2D Anderson -- Hubbard model.}
\label{phd}

\begin{figure}
\includegraphics[clip=true,width=1.0\columnwidth]{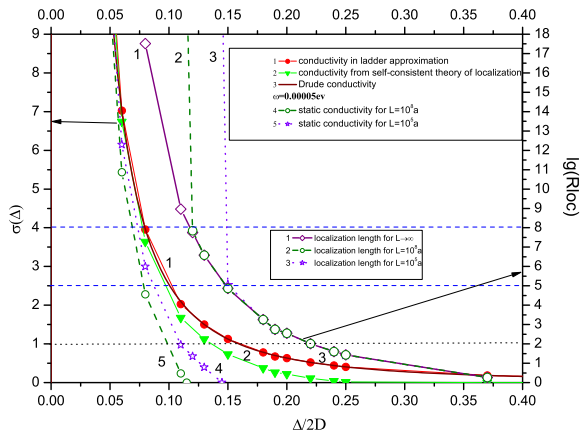}
\caption{Left scale --- dependence of conductivity on disorder $\Delta$
at fixed frequency $\omega=0.00005D$ and $U/2D=1$. Circles -- ladder approximation,
triangles -- self-consistent theory of localization. Curve 3,  
practically coinciding with the results of ladder approximation, is obtained
from Drude expression (\ref{Dr}). Static conductivity for finite samples of sizes 
$L=10^8a$ (curve 4) and $L=10^5a$ (curve 5). Right scale --- dependence of the logarithm
of localization length on disorder $\Delta$: for infinite sample (curve 1) and
for finite samples of sizes $L=10^8a$ (curve 2) and $L=10^5a$ (curve 3).} 
\label{RsRloc} 
\end{figure} 

To proceed further on the left axis of Fig. \ref{RsRloc} we present our data
for the real part of conductivity at fixed and sufficiently small frequency
$\omega=0.00005D$ plotted as a function of disorder strength $\Delta$. 
Circles show results of ``ladder'' approximation, triangles --- results of 
self -- consistent theory of localization (we take $U=0$ here). 
Curve 3, which practically coincides with results of the ``ladder'' 
approximation, is obtained from the usual Drude expression:
\begin{equation} 
\sigma(\omega)=\sigma(0) \frac{\gamma^2}{\gamma^2+\omega^2},
\label{Dr}
\end{equation}
where the static conductivity is given by
$\sigma(0)=e^2N(0)D_0\approx \frac{e^2}{\hbar}\frac{\varepsilon_F}{2\pi\gamma}$, 
with $N(0)$ being the density of states at the Fermi level, $D_0$ is the 
classical diffusion coefficient, $\varepsilon_F\approx D$ is Fermi energy.
Impurity scattering rate was taken as
$\gamma=\pi N(0)\Delta^2\approx \frac{\pi}{2D}\Delta^2$. It can be seen that
the noticeable contribution of localization corrections to conductivity
(clear difference between curve 2 curves 1 and 3) appears only after conductivity
drops below the values of the order of ``minimal'' metallic conductivity
$\sigma_0=\frac{e^2}{\hbar}$ (our data for conductivity are actually normalized 
by this value in all figures). We shall see below that precisely in this region
a kind of Anderson MIT (divergence of localization length) takes place in 2D
systems of reasonable finite sizes.

On the right axis in Fig. \ref{RsRloc} we show our data for the logarithm of
localization length calculated from Eq. (\ref{Rl2}) as a function of disorder
for infinite sample (curve 1) and for finite samples with
$L=10^{8}a$ and $L=10^{5}a$ (curves 2 and 3). It is clearly seen that 
localization length grows exponentially as disorder drops but remains finite
in infinite 2D sample, signifying the absence of Anderson transition. In finite
samples localization length {\em diverges} at some critical disorder (depending
on the system size) demonstrating the existence of an effective Anderson
transition. From Fig. \ref{RsRloc} it is seen that this critical disorder is
achieved when localization length of an infinite system becomes comparable with
characteristic size of the sample: $R_{loc}^{L\to\infty}\sim L$.
It should be noted that in our approach, opposite to the results of 
Ref.~\onlinecite{Shapiro08}, localization length is practically independent of $U$, 
which leads to independence of critical disorder in 2D of correlation strength
$U$. Similar result 
\footnote{Calculations of localization length for 3D system performed by us 
after the publication of Ref.~\onlinecite{HubDis} have demonstrated its practical
independence on the value of $U$}
was obtained in our approach for 3D systems \cite{HubDis}.

On the left axis of Fig.~\ref{RsRloc} disorder dependence of
static conductivity for finite samples of the sizes $L=10^8a$ and $L=10^5a$
(curves 4 and 5) is shown. For finite systems with small disorder 
static conductivity is not zero (metal).
It gradualy goes down while disorder grows and becomes zero at
the same critical value where localization radius diverges on the approach
from insulating phase in a finite sample. 
Static conductivity of finite samples in our calculations
practicaly does not depend on correlation strength $U$.
Rather significant difference between the values of static conductivity
and that of conductivity at small but non zero frquencies seen in Fig.~\ref{RsRloc}
comes from exponential smallness of frequency range of localization
behavior mentioned above.

\begin{figure}[t]
\includegraphics[clip=true,width=1.0\columnwidth]{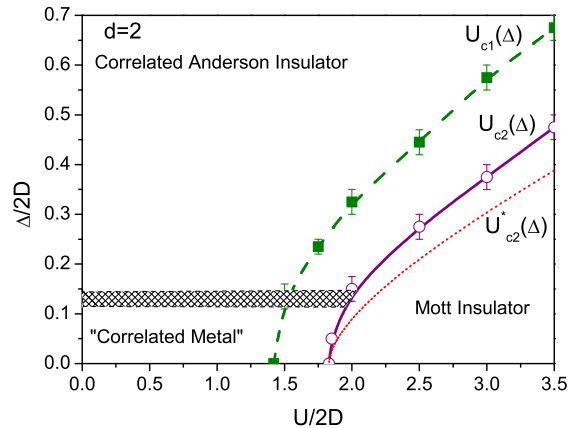}
\caption{Phase diagram of two -- dimensional paramagnetic
Anderson -- Hubbard model at zero temperature. The border of Mott insulator
region $U_{c2}(\Delta)$ and the border of coexistence (hysteresis) region 
$U_{c1}(\Delta)$ are obtained from the evolution of the density of states,
$U^{\ast}_{c2}(\Delta)$ is calculated from (\ref{Ucteor}). Dashed stripe
corresponds to the narrow region, where Anderson metal -- insulator
transition takes place in finite systems.}
\label{ph_diag} 
\end{figure} 

Let us now discuss our results for the phase diagram of 2D half -- filled
Anderson -- Hubbard model, obtained from extensive DMFT+$\Sigma$ calculations of 
DOS and analysis of localization length behavior in finite 2D systems.
The general form of this phase diagram in disorder -- correlation ($\Delta,U$) 
plane is shown in Fig.\ref{ph_diag}.

Dashed stripe in Fig. \ref{ph_diag} corresponds to the region of an effective
transition from Anderson insulator to ``metallic'' phase. It boundaries were
determined by divergence of localization length in finite samples with
characteristic sizes $L=10^{5}a$ (upper boundary) and $L=10^{8}a$ (lower
boundary) (cf. Fig.\ref{RsRloc}). It should be stressed that further increase
of system size, e.g. 10 times up to $L=10^{9}a$, leads only to practically
negligible downshift of the lower boundary (decrease of critical disorder) of 
dashed stripe in Fig. \ref{ph_diag}.

The dependence of $U_{c2}(\Delta)$, obtained from DOS behavior, determines the
boundary for Mott transition and is defined by the disappearance of the
quasiparticle peak in DOS and correlation gap opening at the Fermi level
(cf. Fig.\ref{DOSU},\ref{insDOS}). 

In our previous work \cite{HubDis} on 3D Anderson -- Hubbard model we have 
proposed a simple explanation of $U_{c1,c2}(\Delta )$ dependence. Assuming that the
controlling parameter of Mott -- Hubbard transition given by the ratio of 
Hubbard interaction and effective bandwidth (depending on disorder)
$\frac{U_{c1,c2}(\Delta)}{W_{eff}(\Delta)}$ is universal constant (independent of
disorder), we get: 
\begin{equation}
\frac{U_{c1,c2}(\Delta)}{W_{eff}(\Delta)}=\frac{U_{c1,c2}(0)}{W},
\label{UcW}
\end{equation} 
where $W_{eff}(\Delta)$ is an effective bandwidth in the presence of disorder,
calculated at $U=0$ in self -- consistent Born approximation (\ref{BornSigma}).
In 3D model \cite{HubDis} the dependence of critical correlation strength
on disorder $U_{c1,c2}(\Delta)$, obtained directly from the evolution of DOS, has
shown quite satisfactory agreement with qualitative dependence obtained from
Eq. (\ref{UcW})\footnote{Further extensive calculations performed after the
completion of Ref.~\onlinecite{HubDis} have confirmed practically ideal agreement
between these dependencies.\vspace{0.3cm}}

In 2D model under consideration here, solution of Eq. (\ref{UcW}) gives:
$$U^{\ast}_{c1,c2}(\Delta)=U_{c1,c2}(0)\frac{W_{eff}(\Delta)}{W}=$$
\begin{equation}
=U_{c1,c2}(0)\left(\frac{2{\Delta}^2}{W^2}ln\left(\frac{c+1}{c-1}\right)+c\right),
\label{Ucteor}
\end{equation}
where $c=\sqrt{4\left(\frac{\Delta}{W}\right)^2+1}$. However, unlike in 3D case
\cite{HubDis}, $U_{c2}(\Delta)$ dependence, obtained from DOS evolution is clearly
different from the qualitative one $U^{\ast}_{c2}(\Delta)$ (dotted line in Fig. 
\ref{ph_diag}), determined by Eq. (\ref{Ucteor}). Probably, this is due to a
significant change in the rectangular form of ``bare''DOS with the 
growth of disorder $\Delta$, which is absent for semi -- elliptic ``bare'' DOS 
used in 3D case in Ref.~\onlinecite{HubDis}.

As we already noted, with decrease of $U$  from insulating phase Mott transition 
occurs at $U=U_{c1}(\Delta)<U_{c2}(\Delta)$ and the coexistence (hysteresis) region 
is observed between $U_{c1}(\Delta)$ and $U_{c2}(\Delta)$ curves on phase diagramm 
Fig.\ref{ph_diag}.

\section{Conclusion}
\label{concl}

We have used the generalized DMFT+$\Sigma$ approach to study basic properties
of disordered and correlated Anderson -- Hubbard model. Our method produces
relatively simple interpolation scheme between two well studied limits --- 
that of strongly correlated Hubbard model in the absence of disorder
(DMFT and Mott -- Hubbard MIT) and the case of 2D Anderson insulator in the
infinite system without electron -- electron interactions. It seems that the
proposed interpolation scheme reflects all the qualitative features of 
Anderson -- Hubbard model, such as behavior of the density of states and
dynamic conductivity. The general structure of the phase diagram obtained
in DMFT+$\Sigma$ approximation is also in reasonable agreement with the results 
of direct numerical simulations \cite{PB09}. At the same time, DMFT+$\Sigma$
approach is rather competitive in a sense of the amount of numerical work and
allows direct calculations of all the basic observable characteristics of 
Anderson -- Hubbard model.

It should be stressed that an effective Anderson transition obtained here in the
case of finite size 2D systems is in no sense attributed to electronic 
correlations and follows directly from self -- consistent theory of localization
also in the absence of correlations.  

The main shortcoming of the method used is the neglect of the interference
between disorder scattering and Hubbard interaction, which leads to the
independence of localization length and critical disorder $\Delta_c$ (in finite
2D systems) of correlation strength $U$. The importance of this kind of 
interference effects is known long ago \cite{Lee85,ma}, though these can be 
taken into account only in the case of weak correlations and disorder. At the
same time, the neglect of interference effects is the key point of our
DMFT+$\Sigma$ approach, allowing to obtain rather simple and physically clear
interpolation scheme, allowing to analyze the limits of strong correlations and
disorder. 

Another drastic simplification is our assumption of non magnetic (paramagnetic)
nature of the ground state of Anderson -- Hubbard model. The importance of
magnetic (spin) effects in strongly correlated and disordered systems is obvious, 
as well as the importance of competition between different kinds of magnetic
ground states \cite{georges96}. 

Despite these shortcomings, our results seems rather attractive and reliable,
e.g. with respect to strong disorder effects on Mott -- Hubbard transition and
the general form of the phase diagram at $T=0$. Our predictions for the general
behavior of dynamic (optical) conductivity and disorder induced Mott insulator 
to effective ``metal'' transition can be directly compared with existing and
future experiments.

\section{Acknowledgements}

We are grateful to Th. Pruschke for providing us with his effective NRG code.

This work is partly supported by RFBR grant ÐÔÔÈ 08-02-00021 and programs of
fundamental research of the Presidium of RAS ``Quantum physics of condensed 
state'' and that of Physics Department of RAS ``Strongly correlated electrons
in solids'', as well as by the grant of the President of Russian Federation
MK-614.2009.2 (IN).




\end{document}